\documentclass[twocolumn,showpacs,prl]{revtex4}

\usepackage{graphicx}

\usepackage[center]{subfigure}
\begin{document}

\title{Ordering dynamics in Type-II superconductors}

\author{Nicholas Guttenberg}
\affiliation{Department of Physics, University of Illinois at
Urbana-Champaign, Loomis Laboratory, 1110 West Green Street, Urbana, Illinois, 61801-3080.}
\author{Nigel Goldenfeld}
\affiliation{Department of Physics, University of Illinois at
Urbana-Champaign, Loomis Laboratory, 1110 West Green Street, Urbana, Illinois, 61801-3080.}

\begin{abstract}

We use analytic and numerical methods to analyze the dynamics of
vortices following the quench of a Type-II superconductor under the
application of an external magnetic field.  In three dimensions, in the
absence of a field, the spacing between vortices scales with time $t$
with an exponent $\phi=0.414 \pm 0.01$, In a thin sheet of
superconductor, the scaling exponent is $\phi=0.294 \pm 0.01$. When an
external magnetic field $h$ is applied, the vortices are confined with
respect to the length scale of the Abrikosov lattice, leading to a
crossover between the power-law scaling length scale and the lattice
length scale. From this we suggest a one-parameter scaling of
$\dot{r}$ with $h$ and $r$ that is consistent with numerical data.

\end{abstract}

\maketitle

\section{INTRODUCTION}\label{intro}

Systems with continuous symmetries exhibit interesting time-dependent
scaling behavior following a quench, due to the dynamics of topological
defects \cite{TOYO87, MOND90, MOND92, LIU91, FRAH91, BLUN92, BRAY02}.
Physical realizations include superfluid He$^4$ and He$^3$,
Bose-Einstein condensates, liquid crystals, and models of cosmic
strings - defects produced during phase transitions in the early
universe \cite{KIBB76}. In this paper, we investigate the effect of
topological defects in the approach to equilibrium of Type-II
superconductors and examine the consequences to the dynamics of the
presence of a gauge field \cite{LIU91, FRAH91}.

Following a quench from the normal phase, a Type-II superconductor is
threaded by a series of vortex cores (which are line defects) due to
fluctuations that had existed in the normal phase. Over time, these
vortex cores interact through local and long-range forces. In the
absence of an external field, the number of vortices decreases with
time, giving rise to a characteristic length-scale $r(t)$ that coarsens
with time $t$, generically with a power law whose exponent we denote by
$\phi$.

A variety of predictions have been made for the scaling of $r(t)$ by
treating the problem with various reduced models. A first approximation
to the scaling of defect spacing can be found by solving the equations
of motion for a pair of defects. For local interactions between defects
the strongest contribution is from the immediate neighbor, and the
effect of the remainder of the defects is to distribute momentum via
long-wavelength modes throughout the bulk, leading to over-damped
dissipative dynamics.

We can estimate the dynamics in this approximation from the potential
energy of interaction of a defect pair \cite{BRAY02}. This will play a
role in determining how clusters of defects expand, which in turn
determines the time-dependence of such quantities as the total defect
volume, spacing between defects, and defect correlation functions. If
the interaction is purely from the stress induced by the topological
defect in the continuously symmetric field then the potential will be
of the form $V(r)\propto \log(r/\xi)$ for both point defects in 2D and
line defects in 3D, leading to a force of the form $F(r)\propto
(r/\xi)^{-1}$. Here we have introduced a characteristic length scale
$\xi$, which is of order the vortex core size. Similarly, a point defect
in 3D would experience a potential $V(r)\propto (r/\xi)^{-1}$ and a
force proportional to $(r/\xi)^{-2}$.

In overdamped dynamics, the time-derivative $\dot r(t)$ is proportional
to the force of interaction between two defects. For the $r^{-1}$ force
appropriate for line defects in three dimensions, we find that when
$r(t)$ is much greater than the initial inter-vortex distance, $r
\propto t^{\phi}$ with the dynamical exponent $\phi=1/2$.

In a superconductor,  vortices are surrounded by loops of
superconducting currents, matched to a magnetic field which threads the
vortices. As a result, there are magnetic interactions which depend on
the geometry and dimension of the system in addition to the direct
interaction of the topological defects. In the bulk of a superconductor
the field is screened and the force law is exponentially decaying with
the distance, leading to a logarithmic scaling of the inter-vortex
distance with time. In a thin film of superconducting material (which
will be referred to as the 2DF system), where the magnetic flux is
allowed to penetrate into the third dimension, the force law is that of
point interactions in three dimensional space, and results in a power
law scaling of $r(t)$ with an exponent $\phi = 1/3$.

These exponents are close to those observed in experiment and
simulation, but not exactly so. As first emphasized by
Toussaint and Wilczek \cite{TOUS83}, fluctuations in the initial
conditions can lead to a distribution of topological defects that may
be long-lived, depending on the spatial dimension, thus influencing the
long time scaling exponent $\phi$.  In the context of vortices, such
effects \cite{MOND90} lead to the prediction that $\phi=3/8$, with
subsequent refinements \cite{JANG95, LIU97} yielding the predictions
$\phi=3/10$ for the 2DF system and $\phi=3/7$ for the fully 3D case.

The purpose of this paper is to test these predictions and extend them
to the case of non-zero external applied magnetic field. Our
simulations of the 2DF and 3D bulk superconductor systems yield scaling
laws that are consistent with the predictions of \cite{JANG95, LIU97}.
In the presence of an externally applied magnetic field $h$ (where $h$
is in units of the critical field $H_c$), we found
that the vortex dynamics approach predicts a scaling form for $r(t,h)$,
and that our numerical data collapse to a universal curve after the
initial period of vortex annihilation.

\begin{table}[t]
\begin{tabular}{|c|c|c|}
\hline
Citation & System & $\phi$ \\
\hline
Ref \cite{NISH89} & Complex TDLG & $0.375\pm0.03$ \\
Ref \cite{BLUN92} & 3DXY model & $0.44\pm0.01$ \\
Ref \cite{MOND92} & Complex TDLG & $0.45\pm0.01$ \\
Ref \cite{JANG95} & Langevin dynamics & $0.45\pm0.05$ \\
Ref \cite{LIU97} & 2D Monte-Carlo & $0.42\pm0.02$ \\
\hline
\end{tabular}
\caption{Literature predictions for the dynamic scaling exponent $\phi$}
\label{ResultList}
\end{table}

A summary of the various predictions for the dynamic scaling exponent
is given in Table \ref{ResultList}.

This paper is organized as follows.  In section \ref{overdamped} we
give in more detail the argument to derive the scaling laws from vortex
dynamics in all three systems taking into account the effect of the
fields generated by superconducting currents. We describe in section
\ref{initial} an improvement on the vortex dynamic calculations which
has been used to make a more accurate prediction of the vortex scaling
laws \cite{MOND90, JANG95, LIU97}. This method takes into account
vortex-vortex annihilation and gives the observed scaling laws for both
zero and non-zero external magnetic fields. For the latter case, we
suggest a scaling form for the vortex separation $r$ as a function of
$h$ and $t$ such that the data collapse onto one universal curve:
\begin{equation}
a/\lambda\sqrt{h}-r(t)/\lambda=f(t
\exp(-a/\lambda\sqrt{h})),
\label{eqn:collapse}
\end{equation}
where $a=\lambda\sqrt{\Phi_0}$, $\Phi_0$ is the quantized magnetic
flux of a single vortex, and $\lambda$ is the penetration depth of the
magnetic field into the superconducting bulk.  In section
\ref{simulation} we describe the construction of a simulation to test
these scaling laws, and compare in section \ref{results} the results of
simulation with the predicted scaling and magnetic field data collapse.
We observe the scalings predicted by the vortex annihilation models,
and the non-zero external field data is collapsed onto a single curve
by our proposed choice of variables.

\section{OVERDAMPED VORTEX DYNAMICS}
\label{overdamped}

We now proceed to estimate the scaling laws for the inter-vortex
separation by considering the dynamics of a pair of vortices as
representative of the full many-body system-wide behavior. There are
forces between nearby vortices induced both by the effect they have on
the phase of the wavefunction, and by the magnetic flux which threads
them. The interaction through the phase field falls off at a different
rate than the magnetic interaction, and so there are multiple scaling
regimes.

We earlier gave a heuristic argument for the $\phi=1/2$ scaling law
arising from the phase field interactions. Now we proceed to estimate
in more detail the effects of the internal magnetic field of vortices
on their interactions and scaling. If the vortices are restricted to a
two-dimensional film and the magnetic field can extend into the space
above and below the film, then the interaction at large distances is of
the form $ F(r) \propto 1/r^{2} $ \cite{PEAR64}.

For a three-dimensional block of superconducting material, where the
interactions occur entirely within the block, the magnetic field falls
off over a length scale given by the penetration depth $\lambda$. The
energy of interaction between vortex lines in this case
is \cite{TINK04}:
\begin{equation}
E(r)=\frac{\Phi_0^2}{8 \pi ^2 \lambda ^2}K_0\left(\frac{r}{\lambda}\right),
\end{equation}
where $K_0$ is the zeroth-order Hankel function and $\Phi_0$ is the
magnetic flux per vortex.  $K_0$ behaves logarithmically as $r\rightarrow 0$
and as $\exp(-r/\lambda)/ r^{1/2}$ as $r\rightarrow \infty$.  The
corresponding asymptotic forms for the force $F$ are
$F(r) \sim (r/\lambda )^{-1}$ as $ r\rightarrow 0$
and
$F(r)\sim \exp(-r/\lambda )/r^{1/2}$ as $r\rightarrow \infty$.

We separate out the interactions into nearest-neighbor interactions and
distant interactions. The nearest-neighbor interactions dominate the
expansion of the vortex array with time. In the zero-field case, the
distant interactions never become important as it is possible for
vortices to leave the system at the walls, because the total system
pressure is zero. However, a non-zero applied field corresponds to a
net pressure for vortices with the same field direction, and this
constrains the maximum separation between vortices and leads to a
stable vortex lattice at long time.

For the long-range limit, the exponential decay of the force dominates,
and we have:
\begin{equation}
\sqrt{r}e^r + i\sqrt{\pi} \mathrm{erf} (i \sqrt{r}) \propto t
\end{equation}
The first term of this can be approximately inverted as
$r\rightarrow \infty$ to yield
\begin{equation}
r\sim \ln (t)(1-\frac{\ln (\ln (t))}{2\ln (t)+1})
\end{equation}
At long times, this gives $r\propto \ln (t)$.

In the case of a purely two-dimensional system (2D), the results are
the same ($t^{1/2}$ and $\ln (t)$ scalings). However, in the case of a
two-dimensional slab with three-dimensional fields (2DF), we obtain
$r\propto t^{1/3}$.

These predictions are fair heuristic approximations but do not match
closely our observed scaling exponents. For example, in the 2DF case,
rather than $\phi=1/3$, the observed value is $\phi=0.294
\pm 0.01$.
Similarly, the observed value in the 2D system is $0.414 \pm 0.01$,
rather than the predicted $0.5$.  A possible explanation for these
results is that we are not yet in the long-time limit, due to
long-lived influences of the initial conditions.  During the early
stages of the quench, the largest contribution to the evolution of the
vortex distribution is not repulsion between vortices of the same
topological charge, but rather annihilation between opposite vortices.
That process continues into later times, although with reduced
frequency. We will now explore the implications of such arguments.

\section{THE EFFECT OF INITIAL FLUCTUATIONS}
\label{initial}

\subsection{Zero external magnetic field}
\label{zero}

We now consider the scaling of the inter-vortex spacing during the
period of time in which vortex-vortex annihilation is stronger than the
long-range repulsive forces.  The following argument \cite{MOND90}
predicts a scaling exponent $\phi=3/8$ in two dimensions.

In a two-dimensional sheet perpendicular to the vortex cores, the total
topological charge contained within an area, radius $R$, of the sheet
is related to the integral of the gradient of the phase of the order
parameter around the boundary of that area:
\begin{equation}\int q(r)dA = \frac{1}{2\pi} \int \nabla \theta \cdot d\vec{r}\end{equation}
For a mixture of positive and negative vortices, net rotation at the
boundary will be due mostly to boundary charges within a distance $r$
of the boundary:
\begin{equation}\int q(r)dA \approx \int q(r) dA_{boundary} \approx \sqrt{\rho_0 r R}\end{equation}
This relates the total charge fluctuations in the interior:
\begin{equation} \rho(R) = \frac{N}{R^2} \propto \frac{\sqrt{ \rho_0 r R}}{R^2} \end{equation}
which implies
\begin{equation} \rho(R) \propto R^{-3/2} \end{equation}
With diffusive coarsening of domains, the domain radius will scale as
$\sqrt{t}$, giving $\rho(t) \propto t^{-3/4}$. The spacing between
vortices is proportional to $\rho^{-1/2}$, leading to an inter-vortex
scaling of $t^{3/8}$ in 2D. This argument applies equally well in a
purely 2D system and a purely 3D system, but the growth of the domain
size will be different in a 2DF system.

An analysis by Jang, {\it et al.} \cite{JANG95} extends this argument by
considering how the repulsive force between vortices scales with domain
size, thus introducing corrections to the simple assumption of
diffusive scaling used above.

For the case when the inter-vortex force is proportional to $r^{-1}$, the
force on the boundaries of the domain scale as
\begin{equation}F \propto r^{-1} \propto \sqrt{\rho} \propto R^{-3/4}\end{equation}
Assuming overdamped dynamics of the domain size:
\begin{equation} \frac{dR}{dt} \propto R^{-3/4} \end{equation}
we solve for $R$, with the results that $R \propto t^{4/7}$,
$\rho \propto t^{-6/7}$, and $\phi = 3/7$.

A further extension of this argument was proposed by Liu, {\it et al.} \cite{LIU97}
for systems of arbitrary dimension (relative to the defect dimension),
with the general result that
\begin{equation}2\phi = \nu = \frac{d(d+1)}{d^2+2d-1}\end{equation}
We can use the argument due to \cite{JANG95} to determine the vortex-annihilation scaling
in the two-dimensional slab system, in which the domain scaling is
driven by a $r^{-2}$ force. The result of this is that $R\propto t^{2/5}$
and $r\propto t^{3/10}$.

For the exponentially decaying magnetic interaction of the bulk 3D
superconductor, there is expected to be a logarithmic scaling when the
magnetic interaction term dominates the dynamics.

\subsection{External magnetic field}
\label{external}

If the external field is nonzero, there is an equilibrium spacing
between the vortices. To treat this case, we posit that the interaction
with the walls and remainder of the vortex lattice is of the same form
as the magnetic interaction, with a vortex placed at a distance $2R$.
We can make an argument from the vortex dynamics model to determine the
inter-vortex spacing as a function of time. The methods involving
initial fluctuations and vortex annihilation are more difficult to
apply in this case as one does not simply have expanding domains, but
rather the relaxation into a vortex lattice state with nonzero
equilibrium topological charge.

We assume the existence of the lattice, and look at the
dynamics of a single vortex moving towards its equilibrium position
under the influence of the force
\begin{equation}F(r)=F_0(\lambda(e^{-r/\lambda}-e^{-(R-r)/\lambda}))\end{equation}
We obtain:
\begin{equation}r(t)=\frac{R}{2}\pm\lambda \ln
(\frac{1+\alpha(t)}{1-\alpha(t)})\label{eqn:WallModel}\end{equation}
where
\begin{equation}\alpha(t)=e^{-2\beta (t-t_{0}) e^{-R/2\lambda}},\qquad \beta\equiv F_0\lambda\end{equation}
The equilibrium vortex spacing should scale as $R =
a/(\sqrt{h}\lambda)$ where $a\equiv\lambda \sqrt{\Phi_0}$. This is
simply a consequence of distributing the flux through the sample into
discrete vortices each carrying flux $\Phi_0$. We combine this with our
predicted form, with the result that:
\begin{equation}\frac{a}{\sqrt{h}\lambda}-\frac{r}{\lambda}=f(t
e^{-a/\lambda \sqrt{h}})\label{eqn:WallCollapse}\end{equation}
This form neglects effects due to the interaction of vortices through
the order parameter, and so for very low fields it is expected to fail
in the regime in which one observes power-law dynamics. This can be
remedied in principle by adding a term with the proper force law, but
is undesirable as it introduces more adjustable parameters. It is also
expected to fail at times $t<10$ during which the quench dynamics
dominate vortex dynamics, but should be valid at the
$t\rightarrow\infty$ limit as it reproduces the equilibrium vortex spacing.

A different way of handling this problem is to proceed under the
hypothesis that the entire effect of the external field is to introduce a
crossover between two length scales - the prediction from the previous
analysis for the zero field case and the equilibrium vortex spacing.

The scaling in the absence
of a magnetic field defines a length scale $l_{\phi} \propto t^\phi$,
whereas the Abrikosov vortex lattice spacing produced by the externally applied
magnetic field is a second length scale $l_h \propto 1/\sqrt{h}$
relevant as $t\rightarrow\infty$.
The inter-vortex spacing is expected to have the form
\begin{equation}r(t) = t^{\phi}f(t^{-\phi} h^{-1/2})\label{eqn:PLawCollapse}\end{equation}
such that $f(x)\propto x$ when
$x\rightarrow0$ ($t\rightarrow\infty$, $h\neq0$) and
$f(x)\propto$ constant as $x\rightarrow \infty$ ($h\rightarrow0$, $t\neq0$).

\begin{table}[t]
\begin{tabular}{|c|cccc|}
\hline
System & Vort Dyn & Magnetic & Flucts & Measured\\
\hline
2D & $1/2$ & $\ln$ & $3/7$ & -\\
2DF & $1/2$ & $1/3$ & $3/10$ & $0.294\pm0.01$\\
3D & $1/2$ & $\ln$ & $3/7$ & $0.414\pm0.01$,$\ln$\\
\hline
\end{tabular}
\caption{Summary of predictions of vortex scaling and results}
\label{ResultArray}
\end{table}

A summary of our predictions and the results reported below for our
numerical calculations is given in Table \ref{ResultArray}.

\section{SIMULATIONS}
\label{simulation}

We test this predicted form for the inter-vortex separation by
simulating the complex Ginzburg-Landau equations coupled to a vector
potential \cite{LIU91}:
\begin{equation}\gamma \frac{\partial \psi}{\partial t} = \vec{D}^2 \psi + \psi(\alpha -
|\psi|^2)\label{eqn:OrderParamEq}\end{equation}
where the covariant derivative is given by
\begin{equation}\vec{D}\equiv (\vec{\nabla} - i\vec{A})\end{equation}
We must also simultaneously solve Maxwell's equations in the presence
of a spatially varying conductivity. Using Maxwell's
equations, we obtain \cite{DONA04}:
\begin{equation}\frac{\partial \vec{A}}{\partial t} = \nabla^2
\vec{A}+\Im[\psi^\dagger D \psi]\label{eqn:MaxwellEq}\end{equation}
We have chosen the guage $\nabla\cdot\vec{A}=0$ to simplify Eq.
\ref{eqn:OrderParamEq} and Eqn. \ref{eqn:MaxwellEq}. As a consequence, this
must be explicitly satisfied by the numerical method. By using the
link-variable method we can satisfy this constraint automatically. In this method, the
vector potential is stored along the links between grid centers, and
the order parameter is stored at the grid corners. Differential
operators (the covariant derivative in particular) are evaluated as
integrals around loops containing the corner of interest. The form of
the integrals ensures that even if the gauge field contains a
divergence, it has no effect on the numerical evolution. This technique
has been used to simulate both Type-I and Type-II superconductors
\cite{LIU91, FRAH91, KATO93}.

Using the discretization of \cite{DONA04}, the link variable $\vec{U_i}$
is defined as
\begin{equation}\vec{U_i}\equiv e^{i\vec{A}_i \Delta x}\end{equation}
Then, the covariant derivatives in the forward and backward direction are:
\begin{equation}\vec{D}_i^+\psi = \frac{\vec{U}_{i,x}\psi_{x+i}-\psi_x}{\Delta x}\end{equation}
\begin{equation}\vec{D}_i^-\psi = \frac{\psi_x-\vec{U}_{i,x-i}^*\psi_{x-i}}{\Delta x}\end{equation}
Whenever $D^2$ appears in the equations of motion, it is taken as
$\vec{D}^- \vec{D}^+$ to ensure rotational invariance. The covariant
derivative in the superconducting current term is taken to be $D^+$.
This method is equivalent to making local gauge transformations to
remove the gauge field from the covariant derivative, but as a result
introducing local rotations of the phase of $\psi$.

Often the boundary conditions are the most complicated element of such
simulations. We can simplify the boundary conditions by allowing there
to be a thin shell of insulator around our sample (as we do not have
periodic boundary conditions, this is consistent). The boundary
conditions for the insulator are simpler as one does not need to
consider the behavior of the superconducting current.  The latter would
involve the boundary condition that $\nabla _n\psi
=i\vec{A}_n$ \cite{ENOM91}, and because this contains both values of
$\psi$ and $\vec{A}$ it is not very convenient to solve. We proceed to
generate a layer of insulator by having a spatially-varying value of
$\alpha$ in the Ginzburg-Landau equation. The value $\alpha=1$
corresponds to the superconducting phase being stable (at zero field),
whereas the value $\alpha=-1$ corresponds to the insulator being stable.
The order parameter will fall off to zero smoothly on a scale of the order
of the coherence length across the sharp $\alpha$ boundary, and so the
boundary conditions are automatically taken into account.

To apply an external field, we must choose the form of the vector
potential along the boundaries. We pick a vector potential of the form
$\vec{A}=H(-(y-y_{center})/2,(x-x_{center})/2,0)$. This can pose a
problem, because choosing a particular form of the vector potential on
the boundaries is equivalent to a choice of gauge. So as a consequence,
we break the gauge invariance and thus break the translational symmetry
of the system. We have not found an elegant way around this, but the
range at which the effects become significant can be estimated. The
breaking of symmetry occurs because the gradient of the phase of the
order parameter is slaved to the vector potential. The magnitude of the
vector potential varies over space, attaining a minimum at a point
which we choose when we apply the external field. There is a rotation
of the phase along curves surrounding this point. If the wavelength of
the phase rotation is smaller than the size of a grid cell, then
artifacts will appear in the form of preferred locations for vortices
(pinning).

For a field $H$ applied to a sample with cross-sectional area $A$,
there is a flux $\phi =HA$ through the sample. This creates
$\phi /\Phi_0$ vortices, so the phase of the order parameter
rotates by $2\pi HA/\Phi_0$ along the outer boundary of the system,
which has a length $\sqrt{A}$.  From the constraint for the physical
accuracy of our system, that $\partial _x \theta\le \pi / \Delta x$
where $\Delta x$ is the lattice spacing, we obtain:
\begin{equation}\Delta x \le \frac{\Phi_0}{2 H \sqrt{A}}\end{equation}
which constrains our lattice spacing to prevent a failure to resolve
the order parameter at the boundaries. This requires us to have at least
$N$ grid points for a two-dimensional system of edge length $L$,
where $N=4 L^{4} H^{2} / \Phi_0^{2} $. For a three-dimensional system, this
constraint does not exist along the direction of the magnetic field,
only in the plane perpendicular to it, so the cost there is not
necessarily as bad as $O(L^6)$. Also, for sufficiently small applied magnetic
flux, this constraint on the number of computational cells necessary
is weaker than the constraint that we must be able to resolve the vortex
cores: $\Delta x \le r_{core}$.  This gives the following forms for the
number of computational cells in 2 and 3 dimensions ($N_{2D}$ and
$N_{3D}$):
\begin{equation}N_{2D}=\frac{L^2}{r_{core}^2}\end{equation}
\begin{equation}N_{3D}=\frac{L^3}{r_{core}^3}\end{equation}
An $O(L^6)$ cost is simply intractable, so we must limit ourselves to
weak applied fields.

If the system is not in the equilibrium vortex-lattice state---for
example, if all the vortices are in the center of the system, the
worst-case scenario---then at a distance $r$ from the center the phase
must rotate with a rate $\partial _x \theta = HA/\phi_c r$. There will
be significant nonphysical behavior unless $r$ is smaller than the
vortex core size (so that the order parameter goes to zero). Thus we have
$r_{crit}\le r_{core}$, where $r_{crit}=HA \Delta x / \pi \phi_c$. This
is the other mode of failure. In practice, this is not as stringent a
constraint as there will usually be vortices spread evenly throughout
the system during the simulation of a quench, which reduces the
rotation of the phase near the center.

We have considered ways of removing this constraint: for example, by
evolving the gradient and magnitude of the order parameter, but these
techniques introduce other technical problems. Specifically, the equations of
motion of the gradient of the order parameter have singularities at
vortex cores, and the constraint that vortices be quantized must be
administered separately, requiring that the integral of
the gradient of the order parameter around a closed curve is $2\pi n$.

\begin{figure}[t]
\includegraphics[width=\columnwidth]{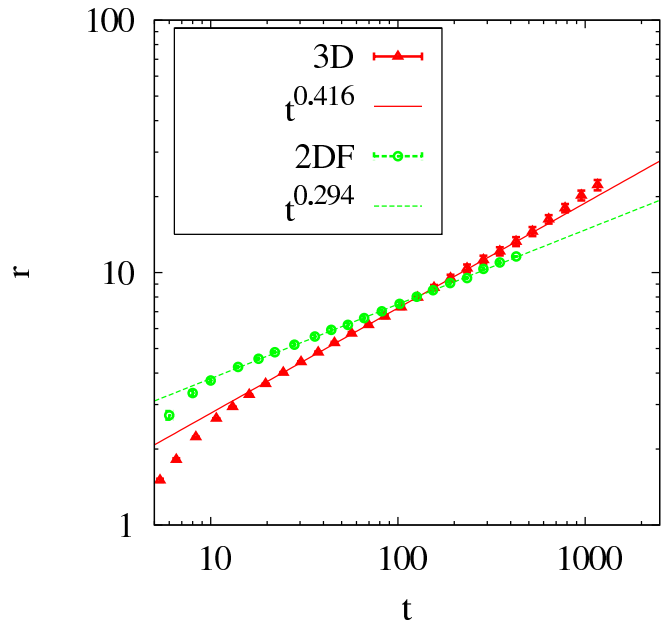}
\caption{ (Color online) This figure shows the power-law scaling of the
2DF ($256\times256\times32$ with $\kappa=1$) and 3D ($92^3$ with
$\kappa=3$) systems at $H=0$ and the deviation from scaling at long times.
The error bars were determined by the standard deviation between different runs. }
\label{H0-plot2}\end{figure}

\begin{figure}[t]
\includegraphics[width=\columnwidth]{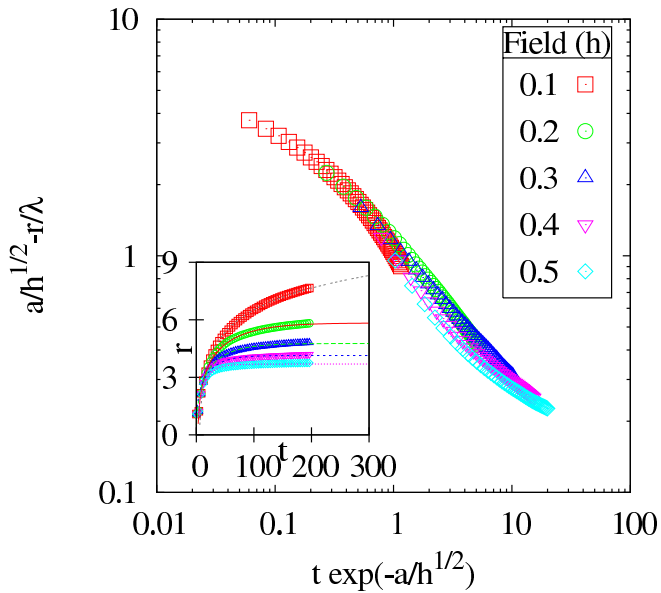}
\caption{ (Color online) Data collapse of the scalings at fields
$h=0.1$ to $h=0.5$. The inset compares the inter-vortex spacing to the
functional form of Eqn. \ref{eqn:WallModel}. The values $a=1.626$ and
$\lambda=1.810$ were used to produce this collapse. These data are from a
3D simulation at size $92^3$ and $\kappa=3$. }\label{Hnz-plot}\end{figure}

\section{RESULTS}
\label{results}

\subsection{Three dimensional superconductor in zero external field}

For the three dimensional block superconductor, we simulated thirty
systems with $96^3$ grid cells with physical dimensions $(32
r_{core})^3$ and $\kappa=3$, starting from a randomized order parameter
and allowed to evolve in time. We also simulated five runs of a $128^3$
system for longer times in order to observe deflections from the
small-$r$ scaling.

The initial state of the order parameter in each cell was chosen by
generating a random angle between $0$ and $2\pi$ and a random amplitude
between $1$ and $0$, both with a flat distribution. In the case where
the superconductor was restricted to a subset of the space (that is, the
2DF system) the amplitude was initialized at zero outside of the thin
plate of superconductor. The three components of the vector potential
were set to random values between $\pm 1\times10^{-6}$ in order to break
the symmetry of the field and reduce potential artifacts.

We compute the average distance between vortices indirectly by
measuring the total volume of vortices in the system and dividing by
the cross-sectional area of a vortex, then using that density to obtain
the inter-vortex spacing:
\begin{equation}r=\sqrt{\frac{\pi r_{core}^2 V}{\int dV (1-|\psi(\vec{r})|^2)}}\end{equation}
Table \ref{ResultArray} summarizes our numerical results and the predictions made
here. For comparison, we provide a list (\ref{ResultList}) of some
previous numerical results that have been obtained in simulations of
similar systems. The majority of such simulations have been done on 2D
and 3D systems with vortex topological defects, but without a gauge
field.

For the 3D bulk superconductor, we observe that in the zero-field case
after about $t=10$, the behavior is power-law, with an
exponent $\phi=0.414\pm0.01$, in comparison with the prediction from
vortex-annihilation dynamics of  $\phi=0.429$.  In the larger
system we observe $t^{0.415}$ scaling, with the behavior deviating from
this around $t=600$ (see Fig. \ref{H0-plot2}), possibly due to finite
size effects.

>From our analysis, we also expect a region of logarithmic scaling.
At long times $r$ varies logarithmically over a time interval
determined by the system size. For all lattice sizes, the behavior
between $t=5$ and $t=20$ seems to follow a logarithmic curve which
then transitions to the power-law behavior.

\begin{figure}[t]
\includegraphics[width=\columnwidth]{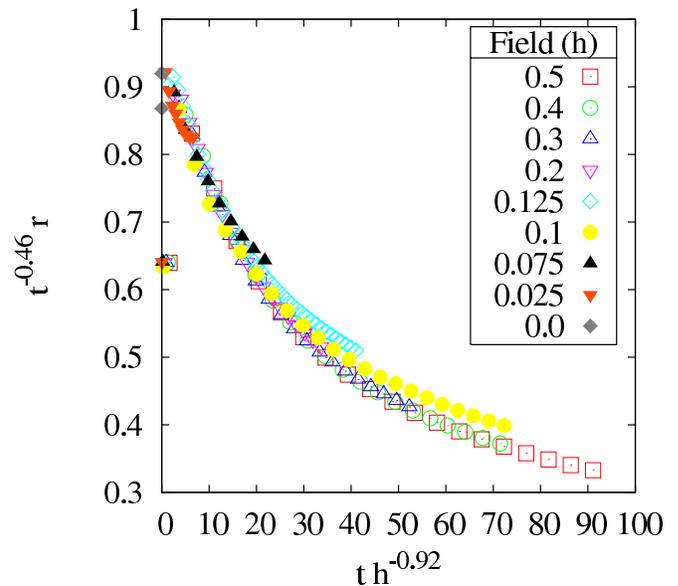}
\caption{ (Color online) Data collapse of the nonzero field scalings
according to Eqn. \ref{eqn:PLawCollapse}. Data are from the same simulation as Fig. \ref{Hnz-plot}. }\label{Datacollapse}
\end{figure}

\subsection{Two-dimensional superconductor}

For the case of a two dimensional sheet with three dimensional fields,
we simulated five runs of a system of size $256\times256\times32$ grid cells, with
$\kappa=1$. The full prediction for that case involves a region of
$t^{3/7}$ scaling, followed by a region of $t^{3/10}$
scaling for vortex-annihilation dominated dynamics. As a consequence,
for intermediate values of $\kappa$ there is no large scaling range
for either regime without waiting long times in the simulation. The smaller
the value of $\kappa$, the shorter the time necessary to observe the
asymptotic regime and the effect of external three-dimensional fields.

We observe a scaling of $t^{0.294\pm0.01}$, which is in agreement with the
prediction for vortex-annihilation dominated scaling in the large-r
limit. The results are shown in Fig. \ref{H0-plot2}.

Eventually, the separation between vortices becomes on the order of the
system size and finite size effects appear in the results which puts the
final limit on the size of the scaling regime we may observe. This
manifests itself as an increase in the fluctuations across different runs and as a
sudden sharp upturn in $r$ compared to the power-law prediction, as
shown by the data for times $t>600$ in Fig. \ref{H0-plot2}.

\subsection{Non-zero field data collapse}

The results for a non-zero applied field are in Fig. \ref{Hnz-plot}.
Here we also plot the best-fit from the predicted form of Eqn.
\ref{eqn:WallModel} along with each curve. The fits are satisfactory,
but break down at short times during the initial quench.

\begin{figure}[t]
\includegraphics[width=\columnwidth]{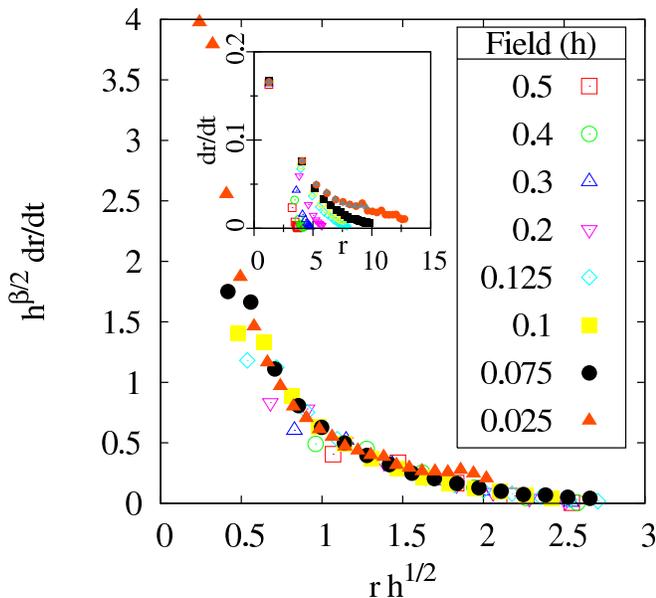}
\caption{(Color online) Data collapse of the nonzero field time derivatives.
$\beta=-1.5$. Data are from the same simulation as Fig. \ref{Hnz-plot}. }\label{Datacollapse2}
\end{figure}

When we plot the data in terms of the combined variables given by Eqn.
\ref{eqn:WallCollapse}, we can observe that the time
and field-strength variables are not independent. The family of curves
we observe will collapse onto a single universal curve given the
proper choice of scaling of the variables as we have described in
section \ref{external}. The behavior in the absence of a magnetic field also has
a crossover - between power-law scaling and logarithmic
scaling. The model which generates this collapse does not take into
account the power-law scaling and so we expect deviations from the
collapse in the case that there is a large regime of power-law scaling
($h\rightarrow0$). Deviations from the collapse
in our simulation are observed at times less than $t=20$ due to the initial
quench behavior which is not captured by this model. At sufficiently
small fields that the Abrikosov lattice spacing is comparable to the
system size we also expect a failure of the collapse at long times -
in the zero field case for the $92^3$ system this occurs at $t>600$.

In order to capture the power-law scaling behavior, we propose a
different form of the data collapse given in Eqn. \ref{eqn:PLawCollapse}.
Plotting the data accordingly we obtain Fig. \ref{Datacollapse}.
The scaling exponent $\phi$ was adjusted to give the
best visual data collapse, which occurred for $\phi=0.46\pm0.2$.

The behavior at short times
does not collapse, which is consistent with the concept behind this scaling: at
short times, the scaling is dominated neither by vortex interactions
nor the external magnetic field. Instead, it is the local dynamics of
the order parameter which contribute to the inter-vortex spacing.
During this time, the measure we have used to determine the inter-vortex
spacing cannot be considered meaningful as there are not yet distinct
vortex cores.

\begin{figure}[t]
\includegraphics[width=\columnwidth]{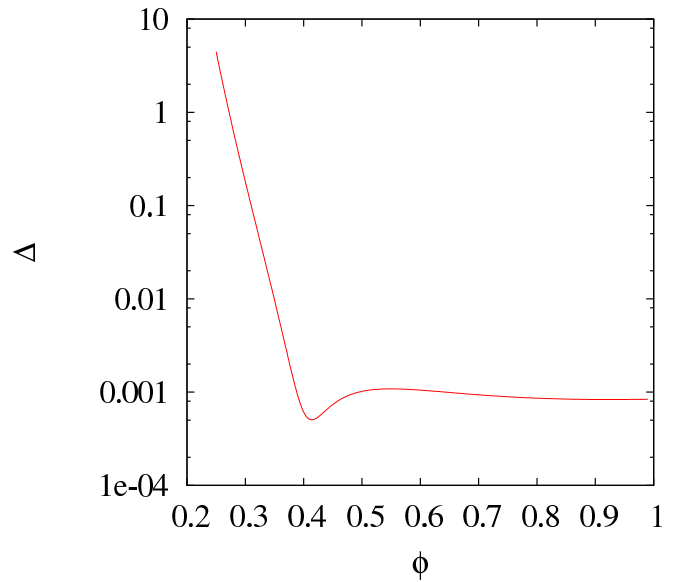}
\caption{(Color online) Error landscape for the $\dot{r}$ data
collapse. }\label{ErrorLandscape}
\end{figure}

We have also examined the time derivative of the vortex spacing,
as this is the basis of the force-law arguments to determine vortex
scaling. The behavior of $\dot{r}$ as a function of $r$
at nonzero external fields asymptotes to zero at
values of $r$ determined by the external field, i.e. the
equilibrium vortex spacings. Rescaling $r$ to map these to the same point
results in a new variable $r^\prime = \sqrt{h}r$. However,
if $\dot{r}$ scales as $r^\beta$ in the limit of small $r$, then
we must also rescale $\dot{r}$ as follows: $\dot{r}^\prime =
h^{\beta/2} \dot{r}$. This collapses the data for $\dot{r}$ onto a
single curve, as seen in Fig. \ref{Datacollapse2}. The value of $\beta$
which best collapses the curves is $\approx -1.5$, which corresponds
to a value of the scaling exponent $\phi=0.4$. Unlike the $r$ data
collapse, the error landscape of the $\dot{r}$ collapse has a
well-defined minimum at finite $\phi$, although there is still a gradual
decrease in the error as $\phi$ becomes unphysically large. Figure
\ref{ErrorLandscape} shows the error in the data collapse measured by
$\Delta\equiv\sum_{ijk} {
(y_j(x_i)-y_k(x_i))^2 }$ where $x_i$ is the $i$th rescaled independent
variable data point and $y_j$ and $y_k$ are the interpolated values of
the rescaled dependant variable for the $j$th and $k$th field strengths.

The data collapse of Fig. \ref{Datacollapse2} is actually an
autonomous counterpart to that of Eqn. \ref{eqn:PLawCollapse} as can be readily
verified by elimination of $t$.

Now we turn to the small field data, which seems not to obey the data
collapse so cleanly. For the $r$ scaling they do not fall onto the data collapse lines nearly
as well as the higher fields. However, for the $\dot{r}$ scaling this
failure of data collapse is not nearly as pronounced. This may be
associated with a bias in the initial spacing introduced by the
period immediately after the quench. That is, for the $r$ scaling it
may be that rather than measuring time from the simulation start, we
must measure it from the end of the quench period when vortex
interactions become the dominant effect. However, for the $\dot{r}$
scaling we avoid this by eliminating the time variable and thus
removing the problem of an arbitrary initial offset.
We will now examine other predictions of the data collapse in order to
understand the small field behavior.

\begin{figure}[t]
\includegraphics[width=\columnwidth]{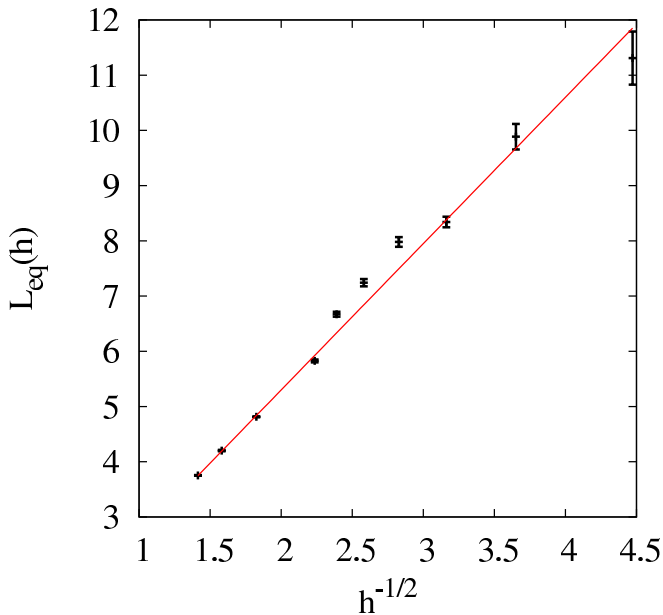}
\caption{(Color online) Equilibrium vortex spacing at nonzero field.
The line is a linear guide.
Data are from the same simulation as Fig. \ref{Hnz-plot}.}\label{AbrikosovCheck}
\end{figure}

The length scale we observe
at long times corresponds to the Abrikosov vortex lattice spacing.
Finite system size effects can potentially produce a crossover to the
system scale and are more likely to do so for weak field values.
The value of the spacing at the longest times accessed by our
simulations is plotted versus the field strength in Fig.
\ref{AbrikosovCheck}. For sufficiently
large fields $h\geq0.2$ (which corresponds to $h^{-1/2}\leq2.24$) we
see a convergence to the predicted scaling within the time range of our simulation.
For small fields we never quite reach the
equilibrium scaling. We have however also looked at a $128^3$ system at
low fields and during the time range of our simulations (up to $t=400$
in this case) we did not observe a significant departure from the $92^3$
curves, so the slight deviation from Abrikosov scaling is more likely to be
a finite time effect than a finite size effect.

\begin{figure}[t]
\includegraphics[width=\columnwidth]{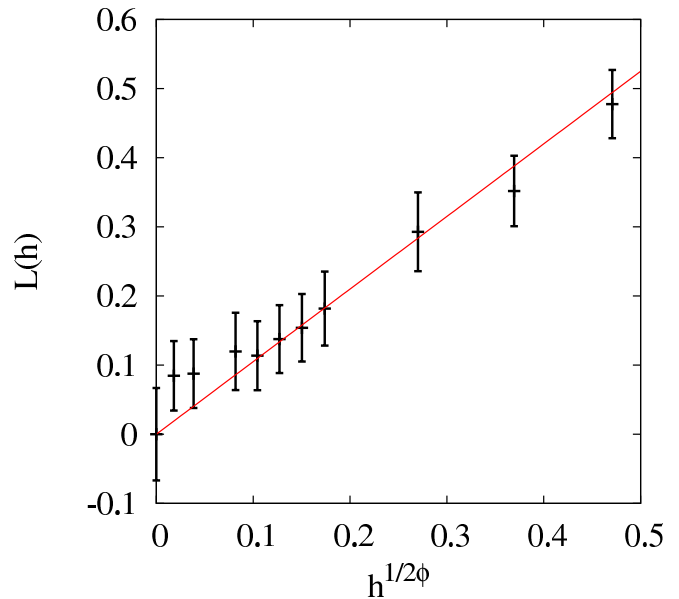}
\caption{(Color online) Scaling of the vortex spacing in the limit of short times.
$\phi=0.46$. The line is a linear guide. Data are from the same simulation as Fig. \ref{Hnz-plot}. }\label{AnalyticPlot}
\end{figure}

We can also make a prediction as to the scaling of $r(h)$ at a fixed
small time $t_1$. Inserting into the scaling form, we get
$r(h,t_1)={t_1}^\phi f({t_1}^{-\phi} h^{-1/2})$. We wish to Taylor expand $f$
around $t=0$, which we may safely do after writing $\bar{f} \equiv
f^{-1/\phi}$, with the result that $r(h,t_1) = {t_1}^{-\phi} \bar{f}(t_1 h^{1/2\phi})
\approx C_1 + C_2 H^{1/2\phi} t_1^{1-\phi} + O({t_1}^{2-\phi})$.

We know that $C_1=0$ due to the required asymptotic behavior of the scaling function, meaning
that we can expect $r(H,t_1) \propto h^{1/2\phi}$ for fixed small
times. The agreement with this prediction is satisfactory, as shown in
Fig. \ref{AnalyticPlot}.

\section{CONCLUSIONS}
\label{conclusions}

We observed both the case of a three dimensional block of
superconductor and a two dimensional sheet via numerical simulation.
In the three dimensional block superconductor, we also examined the
effect of nonzero external fields, and predicted a scaling form based
on the forces between vortices. With this scaling form, we obtained
data collapse to a universal curve independent of the value of the
external magnetic field.
  At zero external field, the dynamical critical exponent $\phi$ was
found to be in agreement with fluctuation/annihilation arguments
rather than the pure vortex-dynamical argument in both the
3D and 2DF systems - that is, in the 3D case we measured
$\phi=0.414\pm0.01$ in comparison with the predicted $3/7$ and
in the 2DF case we measured $\phi=0.294\pm0.01$ in comparison with the
predicted $3/10$. We also observed data suggestive of the logarithmic scaling
regime at short times between the recovery of the order parameter from the
quench and the power-law scaling regime. At long times, we observed
deviation from power-law scaling which appears to be a finite size
effect. We never observed a regime in which all
vortices remaining were of one direction/magnitude of topological
charge in the zero-field case. However, with a nonzero field, the
majority of vortices in the system are aligned with that field, and
so the argument based on vortex dynamics seemed to be more able to predict the
scaling form than in the zero-field cases, in which vortex
annihilation was always a significant contribution.
  We have proposed a form for the behavior of the vortex spacing in
the presence of external magnetic fields which collapses the data onto
a universal curve. We found better collapse when this was applied to
the force law as opposed to the actual vortex spacing likely due to the
initial time offset in which the order parameter recovers from the
quench. This allowed us to explain the behavior at nonzero magnetic field
in terms of a crossover from the zero-field scaling to a fixed length scale
given by the Abrikosov vortex lattice spacing.

\begin{acknowledgments}

We would like to thank Arttu Rajantie, Badri Athreya and Patrick Chan
for helpful discussions. Nicholas Guttenberg was supported by a
University of Illinois Distinguished Fellowship.

\end{acknowledgments}

\bibliographystyle{apsrev}

\bibliography{superconductor}

\end{document}